\documentclass{article}  
\usepackage{amsmath}
\usepackage{amsfonts,amssymb}
\usepackage{epsf}

\newtheorem{theorem}{Theorem}
\newtheorem{corollary}{Corollary}
\newtheorem{lemma}{Lemma}
\newtheorem{remark}{Remark}

\newenvironment{proof}{\noindent\scshape Proof\normalfont.}{\hfill$\Box$}

\newcommand{\const}{\mathrm{const}}
\newcommand{\imag}{i}
\newcommand{\expo}{\mathrm{e}}
\newcommand{\veps}{\varepsilon}
\newcommand{\osym}{o}
\newcommand{\Osym}{O}
\renewcommand{\d}{\mathrm{d}}
\newcommand{\pdr}{\frac{\partial}{\partial r}}
\newcommand{\pdth}{\frac{\partial}{\partial\theta}}
\newcommand{\D}{\mathcal{D}}
\renewcommand{\L}{\mathcal{L}}
\newcommand{\N}{\mathbb{N}}
\newcommand{\Z}{\mathbb{Z}}
\newcommand{\R}{\mathbb{R}}
\newcommand{\C}{\mathbb{C}}

\newcommand{\sign}{\mathrm{sign}\,}
\newcommand{\diag}{\mathrm{diag}\,}
\newcommand{\tr}{\mathrm{tr}\,}

\newcommand{\sdot}{\,\cdot\,}

\begin{document}

\title{{Bound State Solutions of the Dirac Equation \\ in the Extreme Kerr Geometry}}
\author{Harald Schmid \\[1ex] \small NWF I -- Mathematik, Universit{\"a}t Regensburg, \\[-0.5ex]
\small D-93040 Regensburg, Germany, \\[-0.5ex] \small email: \ttfamily harald.schmid@mathematik.uni-regensburg.de}
\date{}
\maketitle

\begin{abstract}
In this paper we consider bound state solutions, i.e., normalizable time-periodic solutions of the Dirac equation in
the exterior region of an extreme Kerr black hole with mass $M$ and angular momentum $J$. It is shown that for each
azimuthal quantum number $k$ and for particular values of $J$ the Dirac equation has a bound state solution, and that
the energy of this Dirac particle is uniquely determined by $\omega = -\frac{kM}{2J}$. Moreover, we prove a necessary
and sufficient condition for the existence of bound states in the extreme Kerr-Newman geometry, and we give an explicit
expression for the radial eigenfunctions in terms of Laguerre polynomials.
\end{abstract}

\section{Introduction}

In Boyer-Lindquist coordinates $(t,\,r,\,\theta,\,\phi)$ with $r\in(0,\infty)$, $\theta\in[0,\pi]$, and
$\phi\in[0,2\pi)$ the metric of a Kerr-Newman black hole of mass $M$, angular momentum $J$, and charge $Q$
is given by (compare \cite[Section 12.3]{Wald})
\begin{equation*}
\d s^{2}
= \frac{\Delta}{U}\left(\d t - a\sin^{2}\theta\,\d\phi\right)^{2}
  - U\left(\frac{\d r^{2}}{\Delta} + \d\theta^{2}\right)
  - \frac{\sin^{2}\theta}{U}\left(a\,\d t - \left(r^{2}+a^{2}\right)\d\phi\right)^{2},
\end{equation*}
where $a := \frac{J}{M}$ is the Kerr parameter and
\begin{equation*}
U(r,\theta) := r^{2}+a^{2}\cos^{2}\theta,\quad
\Delta(r) := r^{2}-2Mr+a^{2}+Q^{2}.
\end{equation*}
In the following we consider the \emph{extreme} case $M^{2} = a^{2}+Q^{2}$, where the function $\Delta$ has only
one zero $\rho := M = \sqrt{a^2+Q^2}$, i.e., $\Delta(r)=(r-\rho)^{2}$. This means, in particular, that the Cauchy horizon
and the event horizon coincide. On such an extreme Kerr-Newman manifold we study the Dirac equation for a particle with
rest mass $m$ and charge $e$ in the exterior region $r\in(\rho,\infty)$. The Dirac equation has the form
\begin{equation} \label{Dirac}
(\mathcal{R}+\mathcal{A})\,\Psi = 0
\end{equation}
with
\begin{align*}
\mathcal{R} & := \left(\begin{array}{cccc}
 \imag mr & 0 & \sqrt{\Delta}\,\D_{+} & 0 \\[1ex]
0 & -\imag mr & 0 & \sqrt{\Delta}\,\D_{-} \\[1ex]
\sqrt{\Delta}\,\D_{-} & 0 & -\imag mr & 0 \\[1ex]
 0 & \sqrt{\Delta}\,\D_{+} & 0 & \imag mr \end{array}\right), \\
\mathcal{A} & := \left(\begin{array}{cccc}
-am\cos\theta & 0 & 0 & \L_{+} \\[1ex]
0 & am\cos\theta & -\L_{-} & 0 \\[1ex]
0 & \L_{+} & -am\cos\theta & 0 \\[1ex]
-\L_{-} & 0 & 0 & am\cos\theta \end{array}\right)
\end{align*}
and the differential operators
\begin{align*}
\D_{\pm} & := \frac{\partial}{\partial r}
              \mp\frac{1}{\Delta}\left[(r^{2}+a^{2})\frac{\partial}{\partial t}
              + a\,\frac{\partial}{\partial\phi}-\imag eQr\right], \\
\L_{\pm} & := \frac{\partial}{\partial\theta} + \frac{\cot\theta}{2}
              \mp\imag\left[a\sin\theta\,\frac{\partial}{\partial t}
              + \frac{1}{\sin\theta}\frac{\partial}{\partial\phi}\right].
\end{align*}
Moreover, by rearranging \eqref{Dirac}, we can write the Dirac equation in Hamiltonian form
\begin{equation} \label{Hamilton}
\imag\,\frac{\partial}{\partial t}\,\Psi = H\Psi,
\end{equation}
where $H$ is a first order $(4\times 4)$ matrix differential operator acting on spinors $\Psi$ on hypersurfaces
$t=\const$. A simple scalar product on such a hypersurface in the exterior region $r\in(\rho,\infty)$ is given by
\begin{equation*}
(\Psi,\Phi) := \int_{\rho}^{\infty}\int_{0}^{\pi}\int_{0}^{2\pi}
\overline{\Psi}(t,r,\theta,\phi)\,\Phi(t,r,\theta,\phi)\sin\theta\,\frac{r^{2}+a^{2}}{\Delta(r)}\,\d\phi\,\d\theta\,\d r,
\end{equation*}
where $\overline{\Psi}$ denotes the complex conjugated, transposed spinor. Note that the Hamiltonian $H$ is in general
not symmetric with respect to this scalar product. However, there exists a scalar product $[\sdot,\sdot]$ on the
spinors on hypersurfaces $t=\const$ which is equivalent to $(\sdot,\sdot)$ such that $H$ is symmetric with respect
to $[\sdot,\sdot]$ (see \cite{FKSY} for the details).

In this paper we are looking for time-periodic solutions
\begin{equation} \label{Periodic}
\Psi(t,r,\theta,\phi)=\expo^{-\imag\omega t}\Psi_0(r,\theta,\phi),\quad\Psi_0\not\equiv 0,
\end{equation}
of the Dirac equation \eqref{Dirac}, where $\omega\in\R$ and $\Psi_0$ is normalizable, i.e.,
$(\Psi_0,\Psi_0)=(\Psi,\Psi)<\infty$. If such a solution exists, then $\omega$ is an eigenvalue of $H$ for the
eigenspinor $\Psi_0$, and $\omega$ represents the one-particle energy of the bound state $\Psi$.
It is well known (see \cite{FSY} and \cite{FKSY}) that normalizable time-periodic solutions do not arise in the
non-extreme case $M^2>a^2+Q^2$ and in the Reissner-Nordstr\o m geometry $a=0$. Here we consider the Dirac
equation on an extreme Kerr-Newman manifold and we prove -- at least for the extreme Kerr case $Q=0$, $a\neq 0$ --
that bound state solutions exist for particular values of $a$. To this end, we employ the ansatz
\eqref{Periodic} with
\begin{equation} \label{Ansatz}
\Psi_0(r,\theta,\phi) = \expo^{-\imag k\phi}
\left(\begin{array}{c}
f_{1}(r)g_{1}(\theta) \\[1ex]
f_{2}(r)g_{2}(\theta) \\[1ex]
f_{2}(r)g_{1}(\theta) \\[1ex]
f_{1}(r)g_{2}(\theta)\end{array}\right),
\end{equation}
where $k\in\{\pm\frac{1}{2},\pm\frac{3}{2},\ldots\}$ is a half-integer, and instead of \eqref{Dirac} we investigate
the equations
\begin{equation*}
\mathcal{R}\Psi =  \lambda\Psi,\quad
\mathcal{A}\Psi = -\lambda\Psi
\end{equation*}
with some separation parameter $\lambda\in\R$. Now, if we define
\begin{equation*}
f(r) := \left(\begin{array}{cc} f_1(r) \\[1ex] f_2(r) \end{array}\right),\quad r\in(\rho,\infty),\qquad
g(\theta) := \left(\begin{array}{cc} g_1(\theta) \\[1ex] g_2(\theta) \end{array}\right),\quad\theta\in[0,\pi],
\end{equation*}
then the Dirac equation can be separated into a radial part
\begin{equation} \label{Radial}
\left(\begin{array}{cc}
(r-\rho)\pdr + \frac{\imag V(r)}{r-\rho} &  \imag m r - \lambda \\[1ex]
-\imag m r - \lambda & (r-\rho)\pdr - \frac{\imag V(r)}{r-\rho}
\end{array}\right)f(r) = 0,
\end{equation}
where $V(r) := \omega\left(r^{2}+a^{2}\right)+k\,a+eQ\,r$, and an angular part
\begin{equation} \label{Angular}
\left(\begin{array}{cc}
\pdth+\frac{\cot\theta}{2}-W(\theta)  & -am\cos\theta+\lambda \\[1ex]
 am\cos\theta+\lambda & -\pdth-\frac{\cot\theta}{2}-W(\theta)
\end{array}\right)g(\theta) = 0,
\end{equation}
where $W(\theta) := a\omega\sin\theta+\frac{k}{\sin\theta}$ (see \cite{Ch}, \cite{Page}).
In the following, a point $\omega\in\R$ is called \emph{energy eigenvalue} of \eqref{Dirac}, if there exist
some $\lambda\in\R$ and nontrivial solutions $f$ of \eqref{Radial}, $g$ of \eqref{Angular} satisfying the
\emph{normalization conditions}
\begin{equation} \label{Integral}
\int_{\rho}^{\infty}|f(r)|^{2}\,\frac{r^{2}+a^{2}}{\Delta(r)}\,\d r
< \infty, \quad \int_{0}^{\pi}|g(\theta)|^{2}\sin\theta\,\d\theta < \infty.
\end{equation}
Then, for the spinor $\Psi$ given by \eqref{Periodic} and \eqref{Ansatz}, it follows that
\begin{equation*}
(\Psi,\Psi) = 2\pi
\left(\int_{\rho}^{\infty}|f(r)|^2\,\frac{r^{2}+a^{2}}{\Delta(r)}\,\d r\right)
\left(\int_{0}^{\pi}|g(\theta)|^2\sin\theta\,\d\theta\right),
\end{equation*}
and the conditions \eqref{Integral} imply $(\Psi,\Psi)<\infty$. This way, the eigenvalue equation $H\Psi_0 = \omega\Psi_0$
and the normalization condition $(\Psi_0,\Psi_0)<\infty$ have been reduced to a pair of boundary value problems for
$(2\times 2)$ systems of ordinary differential equations which are coupled by the energy eigenvalue $\omega$ and the
separation parameter $\lambda$. At first, we try to get some information about the angular eigenvalues $\lambda$ in
dependence of $\omega$. For this purpose, we rewrite in Section 2 the angular Dirac equation \eqref{Angular} and the
second condition in \eqref{Integral} as an eigenvalue equation for some self-adjoint differential operator $A$. From oscillation theory it
follows that $A$ has purely discrete spectrum, and perturbation theory yields that the eigenvalues $\lambda_j$, $j\in\Z$,
of $A$ depend analytically on $am$ and $a\omega$. Subsequently, in Section 3, we investigate the radial Dirac equation
\eqref{Radial} for a fixed azimuthal quantum number $k\in\{\pm\frac{1}{2},\pm\frac{3}{2},\ldots\}$. The first condition
in \eqref{Integral} and the asymptotic behavior of the solutions of \eqref{Radial} at $r=\rho$ and $r=\infty$ imply
\begin{equation*}
\omega = -\frac{ka+eQ\rho}{a^{2}+\rho^{2}}.
\end{equation*}
In this case the system \eqref{Radial} can be reduced either to a Bessel or a Whittaker equation, which allows a more
detailed analysis of the solutions of \eqref{Radial}. By this means, we obtain a necessary and sufficient condition for
$\omega$ being an energy eigenvalue of \eqref{Dirac}. In particular, it turns out that $\omega$ is an energy eigenvalue
if
\begin{equation*}
m^{2}-\omega^{2}>0,\quad\lambda_j^{2}+\rho^{2}m^{2}-\mu^{2}>\frac{1}{4},\quad
n+\frac{\rho m^{2}-\omega\mu}{\sqrt{m^{2}-\omega^{2}}}+\kappa = 0,
\end{equation*}
where $n$ is a positive integer, $j\in\Z\setminus\{0\}$, and
\begin{equation*}
\mu := 2\rho\omega+eQ,\quad\kappa := \sqrt{\lambda^{2}+\rho^{2}m^{2}-\mu^{2}}.
\end{equation*}
Since $\lambda_j$ depends on $\omega$, it is not obvious that these (in)equalities can be satisfied. In order to prove
that bound state solutions of the Dirac equation actually exist, we restrict our attention in Section 4 to the Kerr case
$Q=0$ and $a\neq 0$, where the energy eigenvalue is uniquely determined by $\omega=-\frac{k}{2a}$. Using the estimates
from Section 2, we can show that bound states appear for countably many values of $a$.

\section{The Angular Dirac Equation}

In this section we study the angular part \eqref{Angular} of the separated Dirac equation for some fixed
half-integer $k$. For this reason, we write \eqref{Angular} in the form
\begin{multline} \label{A}
\frac{1}{2\sin\theta}\left[
2\left(\begin{array}{cc} 0 & \sin\theta \\[1ex] -\sin\theta & 0 \end{array}\right)u'(\theta)
+\left(\begin{array}{cc} 0 & \cos\theta \\[1ex] -\cos\theta & 0 \end{array}\right)u(\theta) \right. \\
\left. +\left(\begin{array}{cc}
-2L\sin\theta\cos\theta & 2k+2\Omega\sin^{2}\theta \\[1ex]
 2k+2\Omega\sin^{2}\theta & 2L\sin\theta\cos\theta \end{array}\right)u(\theta)\right]
= \lambda u(\theta),
\end{multline}
where $L:=am$ and $\Omega:=a\omega$. For fixed values of $L$ and $\Omega$, the differential operator $A$ generated by
the left hand side of \eqref{A} is a self-adjoint operator acting on the Hilbert space $L^{2}((0,\pi),2\sin\theta)^{2}$
of square-integrable vector functions with respect to the weight function $2\sin\theta$. From oscillation theory for
Dirac systems (see \cite[Section 16]{Weidmann}) it follows that the spectrum of $A$ consists of discrete eigenvalues
$\lambda_{j}=\lambda_{j}(L,\Omega)$, $j\in\Z$, where $|\lambda_{j}(L,\Omega)|\to\infty$ as $|j|\to\infty$ pointwise on
$\R^2$. Moreover, $A = A(L,\Omega)$ depends analytically on $L$ and $\Omega$, and the partial derivatives with respect
to $L$ and $\Omega$, respectively, are given by
\begin{equation*}
\frac{\partial A}{\partial L} = \left(\begin{array}{cc}
-\cos\theta & 0 \\[1ex] 0 & \cos\theta \end{array}\right),\quad
\frac{\partial A}{\partial\Omega} = \left(\begin{array}{cc}
0 & \sin\theta \\[1ex] \sin\theta & 0 \end{array}\right).
\end{equation*}
Hence, by perturbation theory (see \cite[Chap. VII, \S 3, Sec. 4]{Kato}), also the eigenvalues
$\lambda_{j}=\lambda_{j}(L,\Omega)$, $j\in\Z$, depend analytically on $L$ and $\Omega$, and we obtain the estimates
\begin{equation} \label{L}
\left|\frac{\partial \lambda_{j}}{\partial L}\right|\leq\left\|\frac{\partial A}{\partial L}\right\|\leq 1,\quad
\left|\frac{\partial \lambda_{j}}{\partial\Omega}\right|\leq\left\|\frac{\partial A}{\partial\Omega}\right\|\leq 1,
\end{equation}
where $\|\cdot\|$ denotes the operator norm of a $(2\times 2)$ matrix.

\begin{lemma} \label{Lambda}
The eigenvalues $\lambda_{j}=\lambda_j(L,\Omega)$ of the angular Dirac operator \eqref{A} depend analytically on
$L$ and $\Omega$. Moreover, $|\lambda_{j}|\to\infty$ as $|j|\to\infty$ locally uniformly on $\R^2$.
\end{lemma}

\begin{proof}
For fixed $(L,\Omega)\in\R^2$, we have $|\lambda_{j}(L,\Omega)|\to\infty$ as $|j|\to\infty$, and from
\eqref{L} it follows that $|\frac{\partial \lambda_{j}}{\partial L}|$, $|\frac{\partial\lambda_{j}}{\partial\Omega}|$
are uniformly bounded on $\R^2$.
\end{proof}

\section{The Radial Dirac Equation}

In the following we consider the radial part \eqref{Radial} of the separated Dirac equation in the extreme case
$\Delta(r)=(r-\rho)^{2}$. First, we introduce a new variable $x := r-\rho$ (the coordinate distance from the event
horizon) and we write \eqref{Radial} in the form
\begin{equation} \label{Rx}
f'(x) =
\left(\begin{array}{cc}
-\frac{\imag\tau}{x^{2}}-\frac{\imag\mu}{x}-\imag\omega & \frac{\lambda-\imag m\rho}{x}-\imag m \\[1ex]
 \frac{\lambda+\imag m\rho}{x}+\imag m & \frac{\imag\tau}{x^{2}}+\frac{\imag\mu}{x}+\imag\omega
\end{array}\right)f(x),\quad x\in(0,\infty),
\end{equation}
where
\begin{equation*}
\tau := \omega\left(\rho^{2}+a^{2}\right)+ka+eQ\rho,\quad\mu := 2\rho\omega+eQ.
\end{equation*}
Now, let $S$ be the unitary matrix
\begin{equation} \label{S}
S := \frac{1}{\sqrt{2}}\left(\begin{array}{cc} -1 & \imag\sigma \\[1ex] -\sigma & -\imag \end{array}\right)
\end{equation}
with $\sigma:=\sign\omega$ (and $\sign 0:=1$). By means of the transformation $f(x)=Sw(x)$, the differential equation
\eqref{Rx} is equivalent to the system
\begin{equation} \label{Rw}
w'(x) =
\left(\begin{array}{cc}
\frac{\sigma\lambda}{x} & -\frac{\sigma\tau}{x^{2}}+\frac{\rho m-\sigma\mu}{x}+m-|\omega| \\[1ex]
\frac{\sigma\tau}{x^{2}}+\frac{\rho m+\sigma\mu}{x}+m+|\omega| & -\frac{\sigma\lambda}{x}
\end{array}\right)w(x)
\end{equation}
on the interval $(0,\infty)$, and since $|f(x)|=|w(x)|$, a point $\omega\in\R$ is an energy eigenvalue of \eqref{Dirac}
if and only if \eqref{Rw} has a nontrivial solution $w$ satisfying
\begin{equation} \label{Nw}
\int_{0}^{\infty}|w(x)|^{2}\,\frac{(x+\rho)^{2}+a^{2}}{x^{2}}\,\d x < \infty
\end{equation}
provided that $\lambda$ is an eigenvalue of the angular Dirac operator $A$. In the following we present some
necessary conditions for $\omega\in\R$ being an energy eigenvalue of the Dirac equation \eqref{Dirac}, and we start
with some results on the solutions of singular systems more general than \eqref{Rw}.

\begin{lemma} \label{Oscillation}
Let $y$ be a solution of the differential equation
\begin{equation} \label{D1}
y'(z) = \left(C+R(z)\right)y(z),\quad z\in[1,\infty),
\end{equation}
where $C$ and $R(z)$ are $(2\times 2)$ matrices, $\det C>0$, $R(z)\to 0$ as $z\to\infty$, $R'$ is
integrable on $[1,\infty)$, and $\tr(C+R)\equiv 0$. If $y\neq 0$, then there exists a constant $\delta>0$
such that $|y(z)|\geq\delta$ for all $z\in[1,\infty)$.
\end{lemma}

\begin{proof}
Let
\begin{equation*}
\Lambda := \left(\begin{array}{cc} -\imag\sqrt{\det C} & 0 \\[1ex] 0 & \imag\sqrt{\det C} \end{array}\right).
\end{equation*}
Since $\tr C=0$, $\pm\imag\sqrt{\det C}$ are the eigenvalues of the constant matrix $C$, and there exists an
invertible matrix $T$ such that $T^{-1}CT = \Lambda$. Furthermore, we can fix some point $z_{0}\in[1,\infty)$ such that
$\det\left(C+R(z)\right)>0$ for all $z\in[z_{0},\infty)$. Now, $\pm\imag\sqrt{\det\left(C+R(z)\right)}$ are the
eigenvalues of the matrix $C+R(z)$, and Eastham's Theorem \cite[Theorem 1.6.1]{Eastham} implies that the system
\eqref{D1} has a fundamental matrix $Y(z)=TH(z)\expo^{\imag D(z)}$, where $H(z)\to I$ as $z\to\infty$ ($I$ is the
$(2\times 2)$ unit matrix) and $D$ denotes the diagonal matrix function
\begin{equation*}
D(z) := \diag\left(-\int_{z_{0}}^{z}\sqrt{\det\left(C+R(t)\right)}\,\d t,\int_{z_{0}}^{z}\sqrt{\det\left(C+R(t)\right)}\,\d t\right).
\end{equation*}
If $y$ is a nontrivial solution of \eqref{D1}, then there exists some vector $c\in\C^{2}\setminus\{0\}$ such that
$y(z)=Y(z)c$, and we obtain
\begin{equation*}
c = Y(z)^{-1}y(z) = \expo^{-\imag D(z)}H(z)^{-1}T^{-1}y(z),\quad z\in[z_0,\infty).
\end{equation*}
Since $\expo^{-\imag D(z)}$ is a unitary matrix for all $z\in[z_0,\infty)$, it follows that
\begin{equation*}
|c|=\left|H(z)^{-1}T^{-1}y(z)\right|\leq\left\|H(z)^{-1}T^{-1}\right\||y(z)|,\quad z\in[z_0,\infty).
\end{equation*}
In addition, $\lim_{z\to\infty}H(z)=I$ implies $\liminf_{z\to\infty}|y(z)|\geq c\,\|T^{-1}\|^{-1}>0$. Finally, as $y$ is
continuous on $[1,\infty)$ and $|y(z)|\neq 0$ for all $z\in[1,\infty)$ by the existence and uniqueness theorem, we have
$|y(z)|\geq\delta$ for all $z\in[1,\infty)$ with some constant $\delta>0$.
\end{proof}

\begin{corollary}
If $\omega\in\R$ is an energy eigenvalue of \eqref{Dirac}, then $\tau = 0$. This means,
\begin{equation} \label{E1}
\omega = -\frac{ka+eQ\rho}{a^{2}+\rho^{2}}.
\end{equation}
\end{corollary}

\begin{proof}
Suppose that $\tau\neq 0$, and let $w$ be a nontrivial solution of \eqref{Rw}. By means of the transformation
$y(z)=w\left(\frac{1}{z}\right)$, the differential equation \eqref{Rw} on the interval $(0,1]$ is equivalent to the
asymptotically constant system
\begin{equation} \label{R1}
y'(z) = \left(C+R(z)\right)y(z),\quad z\in[1,\infty),
\end{equation}
where
\begin{equation*}
C := \left(\begin{array}{cc} 0 & \sigma\tau \\[1ex] -\sigma\tau & 0 \end{array}\right),\quad
R(z) := \left(\begin{array}{cc}
-\frac{\sigma\lambda}{z} & -\frac{\rho m-\sigma\mu}{z}-\frac{m-|\omega|}{z^{2}} \\[1ex]
-\frac{\rho m+\sigma\mu}{z}-\frac{m+|\omega|}{z^{2}} & \frac{\sigma\lambda}{z} \end{array}\right).
\end{equation*}
As $\tr C = 0$ and $\det C = \tau^{2}>0$, Lemma \ref{Oscillation} implies that there exists a constant $\delta>0$ such
that $|y(z)|\geq\delta$ on $[1,\infty)$ and therefore $|w(x)|\geq\delta$ on $(0,1]$. Hence, the normalization condition
\eqref{Nw} is not satisfied, and it follows that $\omega$ is not an energy eigenvalue of \eqref{Dirac}.
\end{proof}

Since we intend to find energy eigenvalues of the Dirac equation, we assume in what follows that $\tau=0$ holds. Then
the differential equation \eqref{Rw} becomes
\begin{equation} \label{Sw}
xw'(x) = \left(\begin{array}{cc}
\sigma\lambda &  \rho m-\sigma\mu+(m-|\omega|)x \\[1ex]
\rho m+\sigma\mu+(m+|\omega|)x & -\sigma\lambda \end{array}\right)w(x).
\end{equation}

\begin{lemma} \label{Regular}
Let $y$ be a nontrivial solution of the differential equation
\begin{equation} \label{D2}
xy'(x) = \left(A + xB\right)y(x),\quad x\in(0,1],
\end{equation}
where $A$, $B$ are $(2\times 2)$ matrices and $\tr A=0$, $\det A\geq-\frac{1}{4}$. If $y\neq 0$, then there exists
a constant $\veps>0$ such that $|y(x)|\geq\veps\sqrt{x}$ for all $x\in(0,1]$.
\end{lemma}

\begin{proof}
Let $J$ be the canonical form of $A$. Since $\tr A=0$, $\pm\sqrt{-\det A}$ are the eigenvalues of $A$. Therefore,
\begin{equation} \label{Diagonal}
J = \left(\begin{array}{cc} -\sqrt{-\det A} & 0 \\[1ex] 0 & \sqrt{-\det A} \end{array}\right)
\end{equation}
if $\det A\neq 0$, and we have
\begin{equation} \label{Jordan}
J = \left(\begin{array}{cc} 0 & \nu \\[1ex] 0 & 0 \end{array}\right)
\end{equation}
with some $\nu\in\{0,1\}$ if $\det A = 0$. Moreover, let $T$ be an invertible matrix such that
$T^{-1}AT=J$. By means of the transformation $T\tilde{y}(z)=y\left(\frac{1}{z}\right)$, \eqref{D2} is equivalent to the
system
\begin{equation} \label{D3}
\tilde{y}'(z) = \left(-\frac{1}{z}\,J - \frac{1}{z^2}\,T^{-1}BT\right)\tilde{y}(z),\quad z\in[1,\infty).
\end{equation}
From \cite[Theorem 1.8.1]{Eastham} and the Levinson Theorem (see \cite[Theorem 1.3.1]{Eastham}) it follows that
\eqref{D3} has a fundamental matrix $\tilde{Y}(z)=H(z)z^{-J}$, where $H$ is a continuous $(2\times 2)$ matrix function
which satisfies $\lim_{z\to\infty}H(z)=I$. Hence, $Y(x)=TH\left(\frac{1}{x}\right)x^{J}$ is a fundamental matrix of the
differential equation \eqref{D2}. Now, if $y$ is a nontrivial solution of \eqref{D2}, then there exists some vector
$c\in\C^{2}\setminus\{0\}$ such that $y(x)=Y(x)c$, and we obtain the estimate
\begin{equation} \label{Constant}
|c|=\left|x^{-J}H\left(\textstyle{\frac{1}{x}}\right)^{-1}T^{-1}y(x)\right|\leq b\,\|x^{-J}\|\,|y(x)|,\quad x\in(0,1],
\end{equation}
where $b:=\max_{x\in(0,1]}\|H(\frac{1}{x})^{-1}T^{-1}\|$. If $\det A>0$, then
$x^{-J}=\expo^{-\log x\cdot J}$ is a unitary matrix (since $J$ is a diagonal matrix with purely imaginary entries),
and it follows that $\|x^{-J}\|=1$. In the case that $J$ is the Jordan matrix \eqref{Jordan}, we have
$x^{-J}=I-\log x\cdot J$ and therefore $\|x^{-J}\|\leq 1-\log x\leq \frac{2}{\sqrt{x}}$, $x\in(0,1]$. Finally,
if $-\frac{1}{4}\leq\det A\leq 0$ and $J$ is given by \eqref{Diagonal}, then $\|x^{-J}\|\leq\frac{1}{\sqrt{x}}$. In any
case $\|x^{-J}\|\leq\frac{2}{\sqrt{x}}$, and \eqref{Constant} implies that $|y(x)|\geq\frac{|c|}{2b}\sqrt{x}$ for all
$x\in(0,1]$.
\end{proof}

\begin{corollary} \label{Condition}
If $\omega\in\R$ is an energy eigenvalue of \eqref{Dirac}, then
\begin{equation} \label{E2}
m^{2}-\omega^{2}\geq 0,\quad \lambda^{2}+\rho^{2}m^{2}-\mu^{2}>\frac{1}{4}.
\end{equation}
\end{corollary}

\begin{proof}
Let $w$ be a nontrivial solution of the differential equation \eqref{Sw}. If $m^{2}-\omega^{2}<0$, then Lemma
\ref{Oscillation} implies $|w(x)|\geq\delta$ on $[1,\infty)$ with some constant $\delta>0$, and thus the
normalization condition \eqref{Nw} is not satisfied. Assuming $\lambda^{2}+\rho^{2}m^{2}-\mu^{2}\leq\frac{1}{4}$,
Lemma \ref{Regular} yields $|w(x)|\geq\veps\sqrt{x}$ on the interval $(0,1]$ with some constant $\veps>0$,
and again $w$ does not match the normalization condition \eqref{Nw}.
\end{proof}

In the following we assume that the conditions \eqref{E1} and \eqref{E2} hold. Moreover, we set
\begin{equation*}
\kappa := \sqrt{\lambda^{2}+\rho^{2}m^{2}-\mu^{2}}>\frac{1}{2}.
\end{equation*}

\begin{lemma}
If $\omega\in\R$ is an energy eigenvalue of \eqref{Dirac}, then
\begin{equation} \label{E3}
m^{2}-\omega^{2}>0.
\end{equation}
\end{lemma}

\begin{proof} According to Corollary \ref{Condition}, $\omega$ is not an energy eigenvalue of \eqref{Dirac} if
$m^{2}-\omega^{2}<0$. Now, we suppose that $m=|\omega|$ and we will prove that $\omega$ is not an energy eigenvalue
even in this case. Introducing
\begin{equation*}
w(x) =: \left(\begin{array}{c} u(x) \\[1ex] v(x) \end{array}\right),
\end{equation*}
$\omega$ is an energy eigenvalue of \eqref{Dirac} if and only if the system
\begin{align}
xu'(x) & = \sigma\lambda u(x) + (\rho m-\sigma\mu)v(x), \label{R2a} \\
xv'(x) & = \left[(\rho m+\sigma\mu)+2mx\right]u(x) - \sigma\lambda v(x) \label{R2b}
\end{align}
has a nontrivial solution $(u,v)$ satisfying
\begin{equation} \label{Nu}
\int_{0}^{\infty}\left(|u(x)|^{2}+|v(x)|^{2}\right)\frac{(x+\rho)^{2}+a^{2}}{x^{2}}\,\d x < \infty,
\end{equation}
where $\lambda$ is an eigenvalue of the angular Dirac operator $A$.

If $\rho m-\sigma\mu=0$, then equation \eqref{R2a} implies $u(x)=c_{1}x^{\sigma\lambda}$ with some constant
$c_{1}\in\C$, and from \eqref{Nu} it follows that $c_{1}=0$. Further, from \eqref{R2b} and $u\equiv 0$ we obtain
$v(x)=c_{2}x^{-\sigma\lambda}$ with some constant $c_{2}\in\C$, and \eqref{Nu} gives $c_{2}=0$, i.e., $v\equiv 0$.
Hence, $\omega$ is not an energy eigenvalue of \eqref{Dirac} in the case $\rho m-\sigma\mu=0$.

Next, we assume that $\rho m-\sigma\mu\neq 0$. From \eqref{R2a} it follows that
\begin{equation} \label{R3a}
v(x) = \frac{xu'(x)-\sigma\lambda u(x)}{\rho m-\sigma\mu},
\end{equation}
and replacing $v$ in \eqref{R2b} with \eqref{R3a} gives
\begin{equation} \label{R3b}
x^{2}u''(x)+xu'(x)-\left[\lambda^{2}+\rho^{2}m^{2}-\mu^{2}+2m(\rho m-\sigma\mu)x\right]u(x) = 0.
\end{equation}

We first investigate the case $\rho m-\sigma\mu<0$. By means of the transformation
\begin{equation*}
u(x)=\tilde{u}\left(\sqrt{8m(\sigma\mu-\rho m)x}\right)
\end{equation*}
\eqref{R3b} is on the interval $(0,\infty)$ equivalent to Bessel's differential equation
\begin{equation} \label{B1}
z^{2}\tilde{u}''(z)+z\tilde{u}'(z)+\left[z^{2}-4\kappa^{2}\right]\tilde{u}(z) = 0.
\end{equation}
The Bessel function $J_{\nu}$ and the Neumann function $Y_{\nu}$ of order $\nu=2\kappa\geq 1$ form a fundamental
system of solutions of \eqref{B1}. These functions have the asymptotic behavior
$J_{\nu}(z)\sim\left(\frac{1}{2}z\right)^{\nu}/\Gamma(\nu+1)$,
$Y_{\nu}(z)\sim-\frac{1}{\pi}\Gamma(\nu)\left(\frac{1}{2}z\right)^{-\nu}$
as $z\to 0$ (the properties of the special functions used in this proof and in the following text can be found,
for example, in \cite{AS} or \cite{MOS}). Hence, if $u$ is a solution of \eqref{R3b} which satisfies \eqref{Nu},
then there exists a constant $c\in\C$ such that $u(x)=c\,J_{\nu}(\sqrt{8m(\sigma\mu-\rho m)x})$, and since
$J_{\nu}(z)\sim\sqrt{2/(\pi z)}\cos(z-\frac{2\nu+1}{4}\pi)$ as $z\to\infty$, we obtain $c=0$ according to the
normalization condition \eqref{Nu}. This means, $u\equiv 0$ on $(0,\infty)$, and from \eqref{R3a} it follows that
$v\equiv 0$ on $(0,\infty)$. Therefore, $\omega$ is not an energy eigenvalue of \eqref{Dirac} if $\rho m-\sigma\mu<0$.

Finally, let us consider the case $\rho m-\sigma\mu>0$. By means of the transformation
\begin{equation*}
u(x)=\hat{u}\left(\sqrt{8m(\rho m-\sigma\mu)x}\right),
\end{equation*}
\eqref{R3b} is on the interval $(0,\infty)$ equivalent to the differential equation
\begin{equation} \label{B2}
z^{2}\hat{u}''(z)+z\hat{u}'(z)-\left[z^{2}+4\kappa^{2}\right]\hat{u}(z) = 0.
\end{equation}
The modified Bessel functions $I_{\nu}$ and $K_{\nu}$ of order $\nu=2\kappa\geq 1$, which form a fundamental system
of solutions of \eqref{B2}, asymptotically behave like $I_{\nu}(z)\sim(\frac{1}{2}z)^{\nu}/\Gamma(\nu+1)$ and
$K_{\nu}(z)\sim\frac{1}{2}\Gamma(\nu)(\frac{1}{2}z)^{-\nu}$ as $z\to 0$. Hence, if $u$ is a solution of \eqref{R3b},
then the integrability condition \eqref{Nu} implies $u=c\,I_{\nu}(2\sqrt{8m(\rho m-\sigma\mu)x})$ with some constant
$c\in\C$, and since $I_{\nu}(z)\sim\expo^{z}/\sqrt{2\pi z}$ as $z\to\infty$, it follows that $c=0$. Hence, $u\equiv 0$
on $(0,\infty)$, and \eqref{R3a} gives $v\equiv 0$ on $(0,\infty)$. This proves that $\omega$ is not an energy
eigenvalue of \eqref{Dirac} if $\rho m-\sigma\mu>0$.
\end{proof}

In the following we suppose that the conditions \eqref{E1}, \eqref{E2} and \eqref{E3} are satisfied. Further,
let $T$ be the invertible matrix
\begin{equation} \label{T}
T := \left(\begin{array}{rr}
-\sqrt{m-|\omega|} & \sqrt{m-|\omega|} \\[1ex]
 \sqrt{m+|\omega|} & \sqrt{m+|\omega|} \end{array}\right).
\end{equation}
By means of the transformation $w(x)=Ty(x)$, \eqref{Rw} is on the interval $(0,\infty)$ equivalent to the system
\begin{equation} \label{Ry}
xy'(x) = \left(\begin{array}{cc}
-\alpha-\gamma x & -\beta-\sigma\lambda \\[1ex] \beta-\sigma\lambda & \alpha+\gamma x
\end{array}\right)y(x),
\end{equation}
where
\begin{equation} \label{abc}
\alpha := \frac{\rho m^{2}-\omega\mu}{\sqrt{m^{2}-\omega^{2}}},\quad
\beta  := \frac{(\rho|\omega|-\sigma\mu)m}{\sqrt{m^{2}-\omega^{2}}},\quad
\gamma := \sqrt{m^{2}-\omega^{2}},
\end{equation}
and since $\|T^{-1}\|^{-1}|y(x)|\leq|w(x)|\leq\|T\|\,|y(x)|$, a point $\omega\in\R$ is an energy eigenvalue of
\eqref{Dirac} if and only if the differential equation \eqref{Ry} has a nontrivial solution $y$ satisfying
\begin{equation} \label{Ny}
\int_{0}^{\infty}|y(x)|^{2}\frac{(x+\rho)^{2}+a^{2}}{x^{2}}\,\d x < \infty.
\end{equation}

\begin{theorem} \label{Result}
A point $\omega\in\R$ is an energy eigenvalue of \eqref{Dirac} if and only if there exists an eigenvalue $\lambda\in\R$
of the angular Dirac equation \eqref{Angular} such that
\begin{equation*} \label{E}
\omega = -\frac{ka+eQ\rho}{a^{2}+\rho^{2}},\quad
m^{2}-\omega^{2}>0,\quad\lambda^{2}+\rho^{2}m^{2}-\mu^{2}>\frac{1}{4},
\end{equation*}
and either $\beta-\sigma\lambda=0$, $\alpha+\kappa=0$ or $1+n+\alpha+\kappa=0$ holds with some non-negative integer $n$,
where
\begin{equation*}
\kappa := \sqrt{\lambda^{2}+\rho^{2}m^{2}-\mu^{2}},\quad
\mu := 2\rho\omega+eQ,\quad\sigma := \sign\omega,
\end{equation*}
and $\alpha$, $\beta$, $\gamma$ are given by \eqref{abc}.\\
If $\beta-\sigma\lambda = 0$ and $\alpha+\kappa=0$, then the radial eigenfunctions are constant multiples of
\begin{equation*}
f(x) = x^{\kappa}\expo^{-\gamma x}ST\left(\begin{array}{c}
1 \\[1ex] 0 \end{array}\right),\quad x\in(0,\infty),
\end{equation*}
where the matrices $S$ and $T$ are given by \eqref{S} and \eqref{T}, respectively.\\
If $1+n+\alpha+\kappa=0$, then the radial eigenfunctions are constant multiples of
\begin{equation*}
f(x) = x^{\kappa}\expo^{-\gamma x}ST\left(\begin{array}{c}
(n+1)L_{n+1}^{(2\kappa)}(2\gamma x) \\[1ex]
(\beta-\sigma\lambda)L_{n}^{(2\kappa)}(2\gamma x) \end{array}\right),
\quad x\in(0,\infty),
\end{equation*}
where $L_{n}^{(2\kappa)}$ denotes the generalized Laguerre polynomial of degree $n$ and order $2\kappa$.
\end{theorem}

\begin{proof}
Let $\lambda\in\R$ be an eigenvalue of the angular Dirac equation \eqref{Angular}. Introducing
\begin{equation*}
y(x) =: \left(\begin{array}{c} u(x) \\[1ex] v(x) \end{array}\right),
\end{equation*}
$\omega$ is an energy eigenvalue of \eqref{Dirac} if and only if the system
\begin{align}
xu'(x) & = -(\alpha+\gamma x)u(x) - (\beta+\sigma\lambda)v(x), \label{R4a} \\
xv'(x) & =  (\beta-\sigma\lambda)u(x) + (\alpha+\gamma x)v(x)  \label{R4b}
\end{align}
has a nontrivial solution $(u,v)$ which satisfies
\begin{equation} \label{Nv}
\int_{0}^{\infty}\left(|u(x)|^{2}+|v(x)|^{2}\right)\frac{(x+\rho)^{2}+a^{2}}{x^{2}}\,\d x < \infty.
\end{equation}
We first assume that $\beta-\sigma\lambda=0$. In this case, equation \eqref{R4b} implies
$v(x)=c_{1}x^{\alpha}\expo^{\gamma x}$ with some constant $c_{1}\in\C$, and since $\gamma>0$, \eqref{Nv}
gives $c_{1}=0$. Further, from \eqref{R4a} and $v\equiv 0$ we obtain $u(x)=c_{2}x^{-\alpha}\expo^{-\gamma x}$
with some constant $c_{2}\in\C$, and \eqref{Nv} implies $\alpha < -\frac{1}{2}$ in the case $c_{2}\neq 0$.
A short calculation shows that $\alpha^{2}-\beta^{2}=\rho^{2}m^{2}-\mu^{2}=\kappa^2-\lambda^2$, and therefore
we have $0=\beta^{2}-\lambda^{2}=\kappa^{2}-\alpha^{2}$. Since $\alpha<0<\kappa$, we obtain $\alpha+\kappa=0$.

Now, let $\beta-\sigma\lambda\neq 0$. From \eqref{R4b} it follows that
\begin{equation} \label{U}
u(x) = \frac{xv'(x)-(\alpha+\gamma x)v(x)}{\beta-\sigma\lambda}.
\end{equation}
Replacing $u$ in \eqref{R4a} with \eqref{U} gives
\begin{equation} \label{R5a}
x^{2}v''(x)+xv'(x)-\left[\lambda^{2}+\alpha^{2}-\beta^{2}+(1+2\alpha)\gamma x+\gamma^{2}x^{2}\right]v(x)=0.
\end{equation}
Since $\alpha^{2}-\beta^{2}=\rho^{2}m^{2}-\mu^{2}$ and $\kappa^{2}=\lambda^{2}+\rho^{2}m^{2}-\mu^{2}$, \eqref{R5a}
takes the form
\begin{equation} \label{R5b}
x^{2}v''(x)+xv'(x)-\left[\kappa^{2}+(1+2\alpha)\gamma x+\gamma^{2}x^{2}\right]v(x)=0.
\end{equation}
Further, by means of the transformation
\begin{equation*}
v(x)=\frac{1}{\sqrt{x}}\tilde{v}\left(2\gamma x\right),
\end{equation*}
\eqref{R5b} is equivalent to Whittaker's differential equation
\begin{equation} \label{W}
\tilde{v}''(z)+\left[-\frac{1}{4}-\frac{\frac{1}{2}+\alpha}{z}+\frac{\frac{1}{4}-\kappa^{2}}{z^{2}}\right]\tilde{v}(z)=0.
\end{equation}
A solution of \eqref{W} is the Whittaker function
\begin{equation*}
M_{-\frac{1}{2}-\alpha,\kappa}(z) = z^{\frac{1}{2}+\kappa}\expo^{-\frac{1}{2}z}M(1+\alpha+\kappa,1+2\kappa,z)
\end{equation*}
where $M(p,q,z)$ denotes the Kummer function
\begin{equation*}
M(p,q,z) := \sum_{n=0}^{\infty}\frac{(p)_n}{(q)_n}\frac{z^{n}}{n!}
\end{equation*}
(the Pochhammer symbol is defined by $(p)_n:=p(p+1)\cdots(p+n-1)$ if $n\geq 1$ and $(p)_0:=1$).
Thus, for some constant $c\in\C\setminus\{0\}$,
\begin{equation} \label{Sv}
v(x) = c\,x^{\kappa}\expo^{-\gamma x}M(1+\alpha+\kappa,1+2\kappa,2\gamma x)
\end{equation}
is a nontrivial solution of \eqref{R5b}, and the function $\frac{1}{x}v(x)$ is square integrable in a neighborhood
of $x=0$. Note that $\kappa$ and $-\kappa$ are the characteristic exponents of the differential equation \eqref{R5b}.
Hence, a solution of \eqref{R5b} which is linearly independent of \eqref{Sv} has an asymptotic behavior like
$\left[c_{0}+\osym(1)\right]x^{-\kappa}$ as $x\to 0$ with some constant $c_{0}\neq 0$, and since $\kappa>\frac{1}{2}$,
such a solution cannot satisfy the normalization condition \eqref{Nv}.

Now, $M(p,q,z)=\frac{\Gamma(q)}{\Gamma(p)}z^{p-q}\expo^{z}\left[1+\Osym(1/z)\right]$ as $z\to\infty$ if
$p\neq 0,-1,-2,\ldots,$ and if $n:=-p$ is a non-negative integer, then $M(-n,q,z)$ reduces to a polynomial of degree $n$.
In particular,
$$
M(-n,q,z)=\frac{n!}{(q)_{n}}L_{n}^{(q-1)}(z),
$$
where $L_{n}^{(q-1)}$ denotes the generalized Laguerre polynomial of degree $n$ and order $q-1$. Consequently, if
$-(1+\alpha+\kappa)\not\in\N$, then the solution \eqref{Sv} has the property that
\begin{equation*}
v(x) = c\,x^{\alpha}\expo^{\gamma x}\left[1+\Osym\left(\textstyle{\frac{1}{x}}\right)\right],\quad x\to\infty,
\end{equation*}
with some constant $c\neq 0$, and since $\gamma>0$, $v$ does not match the normalization condition \eqref{Nv}.

In the following we suppose that there exists a non-negative integer $n$ such that $n+1+\alpha+\kappa=0$. In this case,
\begin{align*} \label{V}
v(x) & = (\beta-\sigma\lambda)\frac{(1+2\kappa)_{n}}{n!}x^{\kappa}\expo^{-\gamma x}M(-n,1+2\kappa,2\gamma x) \\
     & = (\beta-\sigma\lambda)x^{\kappa}\expo^{-\gamma x}L_{n}^{(2\kappa)}(2\gamma x)
\end{align*}
is a nontrivial solution of \eqref{R5b}. Applying the differential relation
\begin{equation*}
zM'(p,q,z)+(q-p-z)M(p,q,z) = (q-p)M(p-1,q,z),
\end{equation*}
we can evaluate and simplify the expression in \eqref{U}:
\begin{align*}
u(x) & = \frac{xv'(x)-(\alpha+\gamma x)v(x)}{\beta-\sigma\lambda} \\
     & = (\kappa-\alpha)\frac{(1+2\kappa)_{n}}{n!}x^{\kappa}\expo^{-\gamma x}M(-1-n,1+2\kappa,2\gamma x) \\
     & = (n+1)x^{\kappa}\expo^{-\gamma x}L_{n+1}^{(2\kappa)}(2\gamma x).
\end{align*}
Since $(u,v)$ satisfy the normalization condition \eqref{Nv}, $\omega$ is an energy eigenvalue of the Dirac equation
\eqref{Dirac}, and the radial eigenfunctions are constant multiples of
\begin{equation*}
f(x) = x^{\kappa}\expo^{-\gamma x}ST\left(\begin{array}{c}
(n+1)L_{n+1}^{(2\kappa)}(2\gamma x) \\[1ex]
(\beta-\sigma\lambda)L_{n}^{(2\kappa)}(2\gamma x) \end{array}\right).
\end{equation*}
\end{proof}

As an immediate consequence of Theorem \ref{Result}, we obtain the following well known result (see \cite[Section V]{FSY}).

\begin{corollary} \label{RN}
In the extreme Reissner-Nordstr\o m geometry, bound state solutions of the Dirac equation do not exist.
\end{corollary}

\begin{proof}
In the case $a=0$ we have $\rho=|Q|$, and if a point $\omega\in\R$ is an energy eigenvalue of \eqref{Dirac}, then
$\omega = -e\frac{Q}{|Q|}$ according to Theorem \ref{Result}. The condition $m^2-\omega^2>0$ yields $m^2-e^2>0$,
and since $\mu=2\rho\omega+eQ=-eQ$, we obtain $\alpha=|Q|\sqrt{m^{2}-e^{2}}>0$. Therefore $\alpha+\kappa>\frac{1}{2}$,
and the condition $\alpha+\kappa=-n$ is not satisfied for any non-negative integer $n$. This implies that \eqref{Dirac}
has no bound state solutions of the form \eqref{Ansatz}.
\end{proof}

\begin{remark}
We can also expect that the Dirac equation has no bound state solutions in an extreme Kerr-Newman black hole background
if the angular momentum $J$ is sufficiently small compared to the charge $Q$.
\end{remark}

We conclude this section with the following observation. Applying the unitary transformation
\begin{equation*}
f(x) =: \frac{1}{\sqrt{2}}\left(\begin{array}{rr} 1 & -\imag \\[1ex] 1 & \imag \end{array}\right)
\left(\begin{array}{c} F(x) \\[1ex] G(x) \end{array}\right)
\end{equation*}
and taking into account $\rho=M$, the radial Dirac equation \eqref{Rx} with $\tau=0$ is equivalent to
\begin{align*}
F'(x) & =  \frac{\lambda}{x}\,F(x) - \left[\omega-m-\frac{Mm-\mu}{x}\right]G(x),\\
G'(x) & = -\frac{\lambda}{x}\,G(x) + \left[\omega+m+\frac{Mm+\mu}{x}\right]F(x),
\end{align*}
where $\mu = 2M\omega+eQ$. In Planck units $\hbar=c=G=1$, this system is formally the same as that for the radial Dirac
equation in Minkowski space-time with Coulomb potential $-\frac{\mu}{x}$, Newtonian (scalar) potential $\frac{Mm}{x}$, and
spin-orbit coupling parameter $\lambda$ (see \cite[Example 9.8]{Greiner}).

\section{The Kerr Case}

In this section we consider the Dirac equation on the extreme Kerr manifold, i.e., we assume $Q = 0$ and $M=|a|>0$.

\begin{lemma} \label{Kerr}
If, for some half-integer $k$, the Dirac equation for a particle with rest mass $m$ has a normalizable
time-periodic solution with azimuthal quantum number $k$ in an extreme Kerr geometry with mass $M$ and
angular momentum $J$, then $\frac{|k|}{2}<Mm<\frac{|k|}{\sqrt{2}}$, and the energy of this particle is
given by $\omega = -\frac{kM}{2J}$.
\end{lemma}

\begin{proof}
Let $L:=am$. In the case $Q = 0$, we have $\rho=M=|a|$ and $\mu=2|a|\omega$. Hence, Theorem \ref{Result}
implies that a point $\omega\in\R$ is an energy eigenvalue of \eqref{Dirac} if and only if for some $j\in\Z$
\begin{equation} \label{K}
a\omega = -\frac{k}{2},\quad |a\omega|<L,\quad\lambda_j(L)^2 + L^{2} - k^{2}>\frac{1}{4},
\end{equation}
and either
\begin{equation*}
\beta(L)-\sign(\omega)\lambda_{j}(L)=0,\quad\alpha(L)+\kappa_j(L)=0
\end{equation*}
or
\begin{equation*}
1+n+\alpha(L)+\kappa_j(L)=0
\end{equation*}
with some non-negative integer $n$, where $\lambda_{j}(L)$ denotes the $j$-th eigenvalue of the angular Dirac
equation \eqref{A} with $\Omega = -\frac{k}{2}$, and
\begin{equation*}
\alpha(L) := \frac{2L^2-k^2}{\sqrt{4L^2-k^2}},\quad
\beta(L)  := -\frac{|kL|}{\sqrt{4L^2-k^2}},\quad
\kappa_{j}(L) := \sqrt{\lambda_{j}(L)^{2}+L^2-k^2}.
\end{equation*}
A necessary condition for $\omega=-\frac{k}{2a}=-\frac{kM}{2J}$ being an energy eigenvalue of the Dirac equation
\eqref{Dirac} is $\alpha(L)<0$, i.e., $\frac{k^2}{4}<L^2<\frac{k^2}{2}$, where $L^2=a^2m^2=M^2m^2$.
\end{proof}

Since $k$ is a half-integer, we have $|k|\geq\frac{1}{2}$, and Lemma \ref{Kerr} immediately yields the
following result.

\begin{corollary}
The inequality
\begin{equation*}
Mm>\frac{1}{4}
\end{equation*}
is a necessary condition for the existence of bound states in the extreme Kerr geometry. Here, $M$ is the mass of the
extreme Kerr black hole and $m$ is the rest mass of the Dirac particle.
\end{corollary}

\begin{theorem}
For a fixed half-integer $k$, there exist two sequences $(a_{n}^{-})_{n\in\N}$ and $(a_{n}^{+})_{n\in\N}$
with the properties
\begin{equation*}
a_n^{-}<0<a_n^{+},\quad |a_n^{\pm}|\in\left(\frac{|k|}{2\,m},\frac{|k|}{\sqrt{2}\,m}\right),
\quad\lim_{n\to\infty}|a_n^{\pm}|=\frac{|k|}{2m},
\end{equation*}
such that the Dirac equation has a normalizable time-periodic solution with azimuthal quantum number $k$
in the extreme Kerr black hole background with mass $M=|a_n^{\pm}|$ and angular momentum $J=a_{n}^{\pm}M$,
and the one-particle energy of this bound state is given by $\omega = -\frac{k}{2a_{n}^{\pm}}$.
\end{theorem}

\begin{proof}
Again, let $L:=am$. From Lemma \ref{Kerr} it follows that energy eigenvalues of the Dirac equation appear at most in the
case $|L|\in\left(\frac{|k|}{2},\frac{|k|}{\sqrt{2}}\right)$. Now, by Lemma \ref{Lambda}, the function
$\lambda_{j}$ depends continuously on $L$, and $|\lambda_{j}|\to\infty$ uniformly on the compact set
$K:=\left[-\frac{|k|}{\sqrt{2}},-\frac{|k|}{2}\right]\cup\left[\frac{|k|}{2},\frac{|k|}{\sqrt{2}}\right]$
as $|j|\to\infty$. Hence, there exists an integer $j=j(k)$ such that the last inequality in \eqref{K} holds for all
$L\in K$. Since $\alpha(L)\to -\infty$ as $|L|\to\frac{|k|}{2}$, $\alpha(L)\to 0$ as $|L|\to\frac{|k|}{\sqrt{2}}$,
$\alpha(L)<0$ if $\frac{|k|}{2}<|L|<\frac{|k|}{\sqrt{2}}$, and $\alpha$, $\lambda_j$ are continuous functions on $K$,
the intersection theorem yields that for fixed $n\in\N$ the equation $\alpha(L) = -1-n-\kappa_j(L)$ has at least one
solution $L_n^{-}=a_{n}^{-}m$ in the interval $(-\frac{|k|}{\sqrt{2}},-\frac{|k|}{2})$ and at least one solution
$L_n^{+}=a_{n}^{+}m$ in the interval $(\frac{|k|}{2},\frac{|k|}{\sqrt{2}})$. In addition,
$\alpha(L_n^{\pm}) = -1-n-\kappa_j(L_n^{\pm})\to-\infty$ as $n\to\infty$, which implies
$\lim_{n\to\infty}|a_n^{\pm}|=\frac{|k|}{2m}$.
\end{proof}

\begin{remark}
We can expect that a similar result holds for extreme Kerr-Newman black holes if the charge $Q$ is sufficiently
small compared to the angular momentum $J$.
\end{remark}

\begin{remark}
Since the functions $\alpha$ and $-n-\kappa_j$ both depend analytically on $L$, but are not identical,
there can exist at most finitely many solutions of the equation $\alpha(L) = -n-\kappa_j(L)$ for each
half-integer $k$, $n\in\N_0$, and $j\in\Z$. In particular, bound states for the Dirac equation exist on at most
countably many extreme Kerr black holes.
\end{remark}

The figures below are the results of numerical computations and give some examples for the radial and angular
density functions $|f|^2$, $|g|^2$ of the bound state solutions for different values of $k$ and $a$. In each example,
the half-integer $k$ and the corresponding values of $am$ and $\frac{\omega}{m}$ in Planck units $\hbar=c=G=1$
are specified.

\begin{figure}[ht]
\begin{center}
\hspace*{8ex}\epsfxsize = 0.5\linewidth
\epsfbox{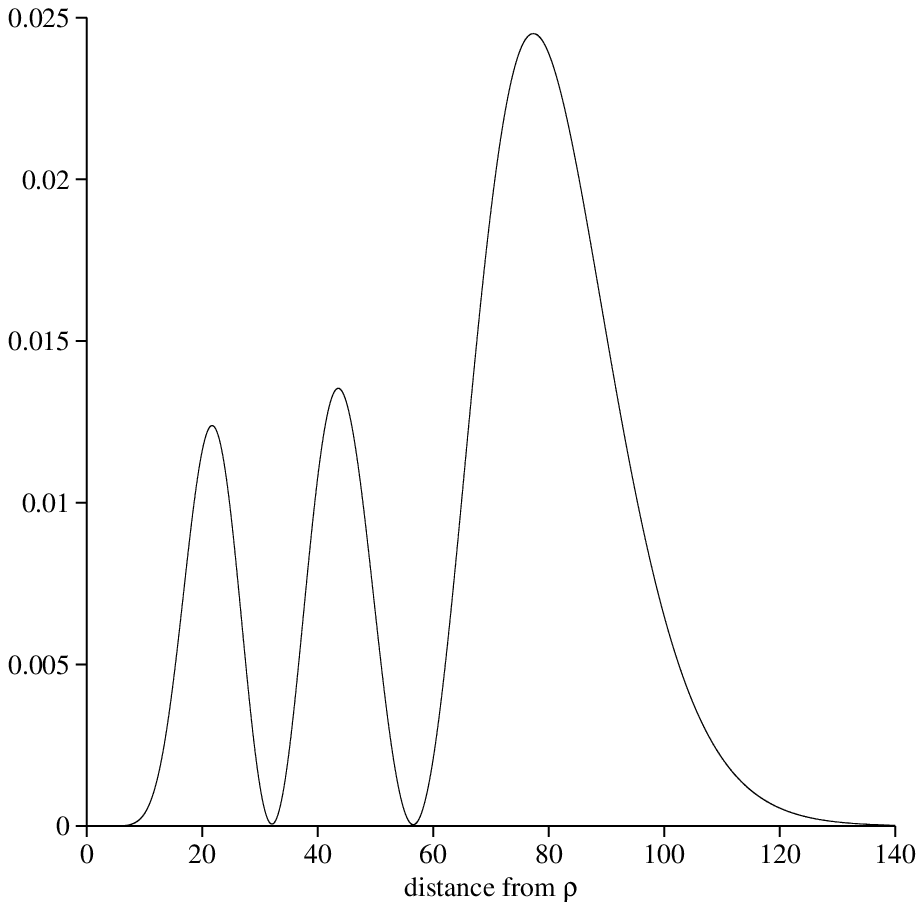}\nobreak
\epsfxsize=0.5\linewidth
\epsfbox{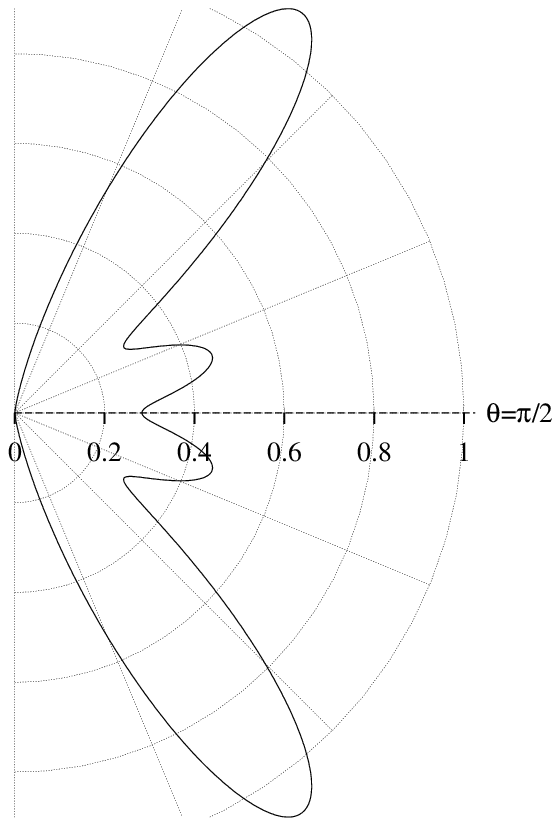}
\end{center}
\caption{$k=\frac{5}{2}$, $am = -1.264065$, $\frac{\omega}{m} = 0.988873$}
\end{figure}
\begin{figure}[ht]
\begin{center}
\hspace*{8ex}
\epsfxsize=0.5\linewidth
\epsfbox{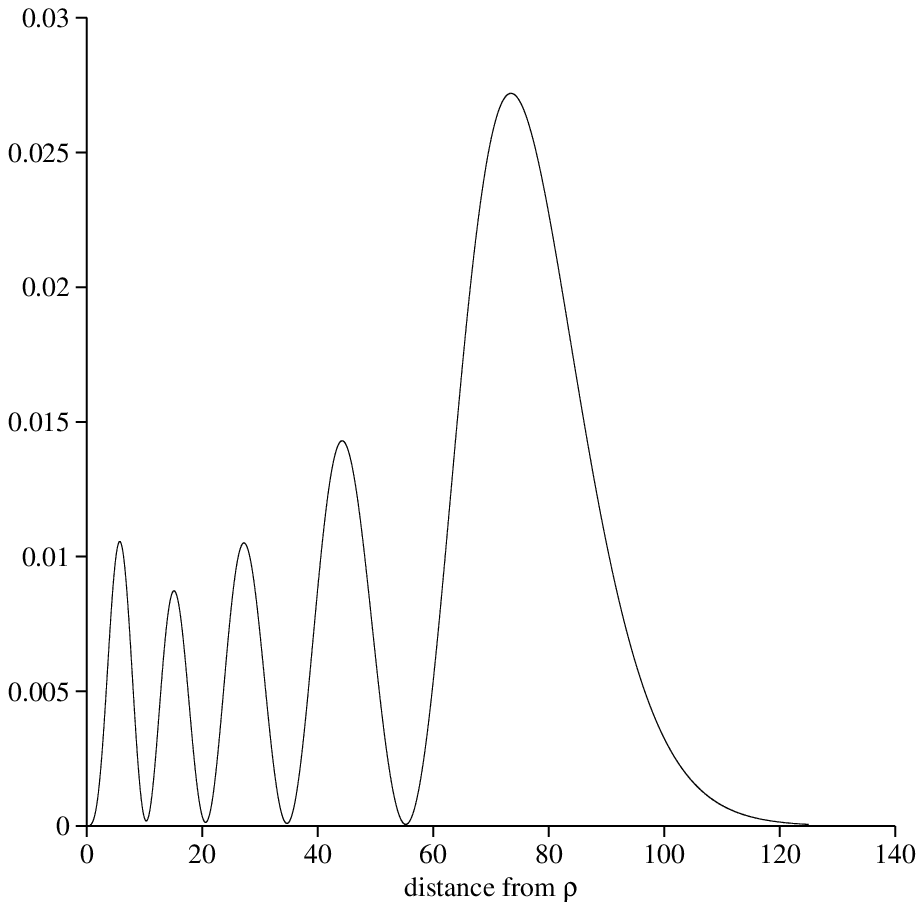}\nobreak
\epsfxsize=0.5\linewidth\epsfbox{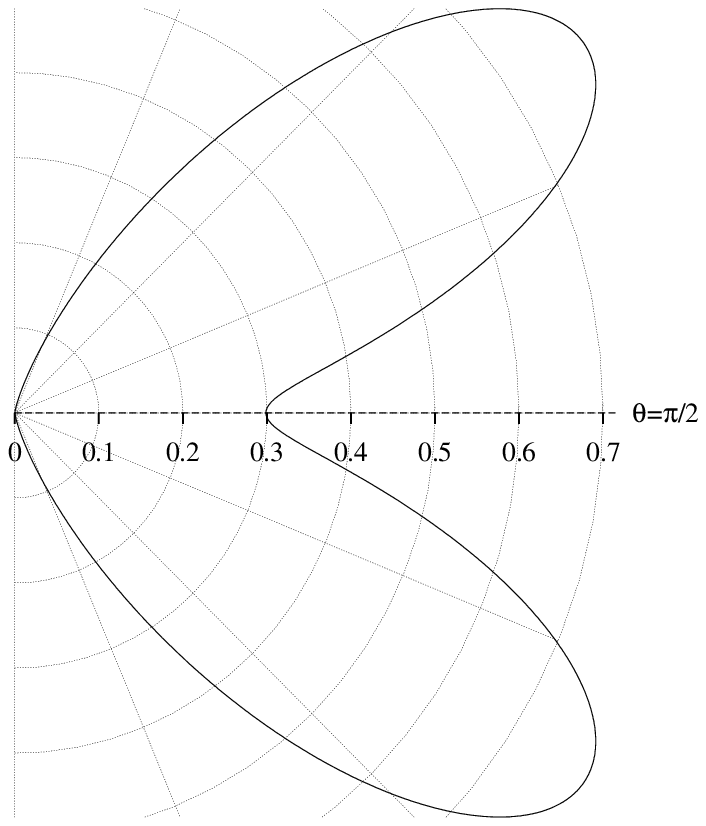}
\end{center}
\caption{$k = \frac{5}{2}$, $am = -1.266630$, $\frac{\omega}{m}=0.986871$}
\end{figure}
\begin{figure}[ht]
\begin{center}
\hspace*{8ex}
\epsfxsize=0.5\linewidth
\epsfbox{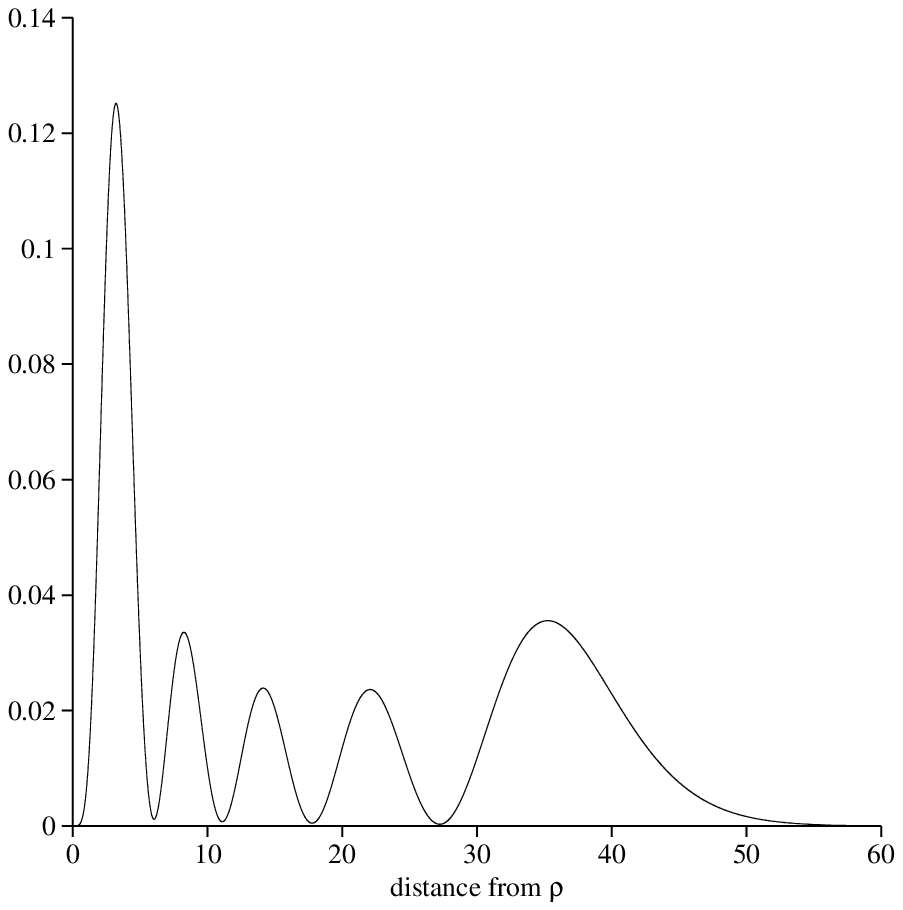}\nobreak
\epsfxsize=0.5\linewidth
\epsfbox{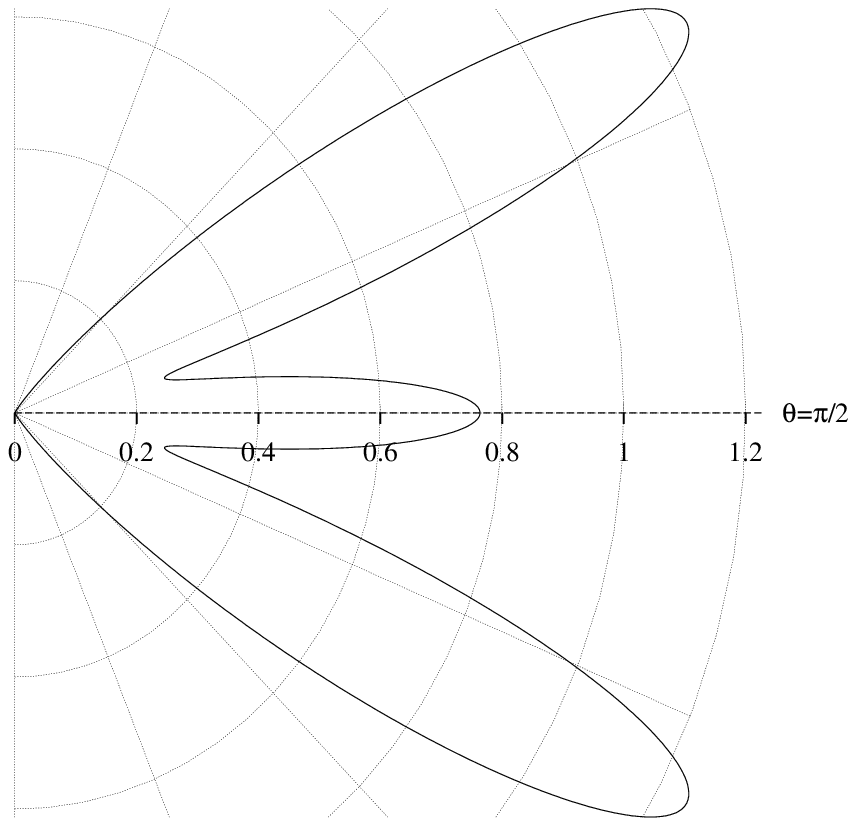}
\end{center}
\caption{$k = \frac{17}{2}$, $am = -4.594167$, $\frac{\omega}{m} = 0.925086$}
\end{figure}

\section*{Acknowledgement}
The author would like to thank Felix Finster, Universit{\"a}t Regensburg, Germany, and Monika Winklmeier,
Universit{\"a}t Bremen, Germany, for helpful suggestions and remarks.


\begin{thebibliography}{12}

\bibitem{AS}
{\sc M.~Abramowitz and I.~A.~Stegun (eds.)},
{\em Handbook of mathematical functions, with formulas, graphs, and mathematical tables},
Dover Publications, Inc., New York, 1966.

\bibitem{Ch}
{\sc S.~Chandrasekhar},
{\em The solution of Dirac's equation in Kerr geometry},
Proc. Roy. Soc. Lond. A {\bf 349} (1976), no. 1659, 571--575.

\bibitem{CL}
{\sc E.~A.~Coddington and N.~Levinson},
{\em Theory of Ordinary Differential Equations},
McGraw Hill Company, Inc., New York -- Toronto -- London, 1955.

\bibitem{Eastham}
{\sc M.~S.~P.~Eastham},
{\em The Asymptotic Solution of Linear Differential Systems. Applications of the Levinson Theorem},
London Mathematical Society Monographs (New Series) 4, Oxford University Press, New York, 1989.

\bibitem{FSY}
{\sc F.~Finster, J.~Smoller, and S.-T.~Yau},
{\em Non-existence of time-periodic solutions of the Dirac equation in a Reissner-Nordstr\o m black hole background},
J. Math. Phys. {\bf 41} (2000), no. 4, 2173--2194.

\bibitem{FKSY}
{\sc F.~Finster, N.~Kamran, J.~Smoller, and S.-T.~Yau},
{\em Nonexistence of time-periodic solutions of the Dirac equation in an axisymmetric black hole geometry},
Commun. Pure Appl. Math. {\bf 53} (2000), no. 7, 902--929.

\bibitem{Greiner}
{\sc W.~Greiner},
{\em Relativistic Quantum Mechanics. Wave Equations},
Theoretical Physics: Text and Exercise Books, 3, Springer, Berlin, 1990.

\bibitem{Kato}
{\sc T.~Kato}, {\em Perturbation theory for linear operators}, Springer, Berlin -- Heidelberg -- New York, 1966.

\bibitem{MOS}
{\sc W.~Magnus, F.~Oberhettinger, R.~P.~Soni},
{\em Formulas and Theorems for the Special Functions of Mathematical Physics},
Springer, Berlin -- Heidelberg -- New York, 1966.

\bibitem{Page}
{\sc D.~Page}, {\em Dirac equation around a charged, rotating black hole}, Phys. Rev. D {\bf 14} (1976), pp. 1509.

\bibitem{Wald}
{\sc R.~M.~Wald}, {\em General Relativity}, The University of Chicago Press, Chicago -- London, 1984.

\bibitem{Weidmann}
{\sc J.~Weidmann}, {\em Spectral Theory of Ordinary Differential Operators},
Lecture Notes in Mathematics {\bf 1258}, Springer, Berlin -- New York, 1987.

\end{thebibliography}
\end{document}